# On-lattice voxelated convolutional neural networks for prediction of phase diagrams and diffusion barriers in cubic alloys

S. Mohadeseh Taheri-Mousavi[1,2,*], S. Sina Moeini-Ardakani[3], Ryan W. Penny[1], Ju Li[2,4,*], A. John Hart[1,*]

*Correspondence to: smousavi@mit.edu (S.M.T.M.), liju@mit.edu (J.L.), ajhart@mit.edu (A.J.H.)

[1]Department of Mechanical Engineering, Massachusetts Institute of Technology, 77 Massachusetts Avenue, Cambridge, MA 02139, USA.
[2]Department of Materials Science and Engineering, Massachusetts Institute of Technology, 77 Massachusetts Avenue, Cambridge, MA 02139, USA.
[3]Department of Civil and Environmental Engineering, Massachusetts Institute of Technology, 77 Massachusetts Avenue, Cambridge, MA 02139, USA.
[4]Department of Nuclear Science and Engineering, Massachusetts Institute of Technology, 77 Massachusetts Avenue, Cambridge, MA 02139, USA.

**Abstract**

Cluster expansion approximates an on-lattice potential with polynomial regression. We show that using a convolutional neural network (CNN) instead leads to more accurate prediction due to the depth of the network. We construct our CNN potential directly on cubic lattice sites, representing voxels in a 3D image, and refer to our method as the voxelated CNN (VCNN). The convolutional layers automatically integrate interaction terms in the regressor; thus, no explicit definition of clusters is required. As a model system, we combine our VCNN potential with Monte Carlo simulations on a $Ni_{1-x}Al_x$ ($x < 30\%$) and predict a disordered-to-ordered phase transition with less than 1 meV/atom error. We also predict the energetic landscape of vacancy diffusion. Classification of formation energy with respect to short-range-ordering of Al alloys around a vacancy reveals that the ordering decreases the probability of Ni diffusion. As the width of our input layer does not depend on the atomic composition, VCNNs can be applied to study alloys with arbitrary numbers of elements and empty lattice sites, without additional computational costs.

**Introduction**

On-lattice techniques such as the Ising model and cluster expansion are widely used to predict phase transitions in a variety of material systems[1–3]. Similar to the Feynman diagrams for many-body systems, the main logic behind these on-lattice models is construction of a regressor for the prediction of a target energy, by the summation of self, binary, ternary, and higher-order interaction terms. This process is also similar to constructing a single layer neural network (NN) with linear addition of a variety of interaction terms. However, the number of interaction terms explodes combinatorially as the number of species increases, resulting in an extremely wide single layer. Here, we show that using on-lattice convolutional NN (CNN) potentials represents a versatile approach to predict target properties by benefiting from the universal synergy between architecture depth and accuracy in NN interpolations[4].

Our generalized on-lattice CNN potential for metal alloys is directly built on lattice sites. These sites represent voxels in a 3D image with a distinct color indicating their types (Fig. 1), which we refer to as a voxelated CNN (VCNN). From such images, through layers of the CNN architecture, convolutional filters automatically extract the interactions among alloying elements and enable identification of critical local atomic arrangements that influence properties. This is analogous to image analysis with deep learning architectures in computer vision.[5] Therefore, the automatic extraction circumvents all preprocessing steps required in the manual specification of these interactions in on-lattice clustering and Ising models. The VCNN method thus can be generalized to multi-component alloys with arbitrary numbers of elements and with empty lattice sites, without additional computational costs, as the input CNN layer has a fixed width. Further, the inherent nonlinear interpolation of VCNN regression by increasing the depth of the architecture improves the prediction accuracy. We applied the VCNN to predict the finite-temperature phase diagram of a face-centered-cubic (fcc) $Ni_{1-x}Al_x$ ($x < 30$ at. %) model system with less than 1 meV/atom error. We also predicted attributed energies for bulk diffusion mediated by exchanging sites with a vacancy in $Ni_{1-x}Al_x$ ($x = 15$ at. %). We analyzed the influence of Al atoms' propensity to form short-range order (SRO) around a vacancy on these energies to exemplify the effect of a local chemical environment on global material properties.

**Results and discussion**

First, the input tensor containing numerical indices representing the types of the elements in all potential lattice sites is built; each lattice site may be occupied by an alloying element (1=Ni, 2=Al, etc.) or remain empty (0=vacancy/empty site). The latter case also includes vacancies in the case of metallic alloys (Fig. 1B). CNNs containing different numbers of convolutional (CONV) and fully connected (FC) layers are built on this image representation (Fig. 1C). CONV layers apply filters on the neighboring voxels of the input tensors and update the output feature set vectors after applying rectified linear unit activation functions. The feature set vectors then pass through the FC layers with nonlinear functions. The output layer computes the overall feature set vector containing the predicted global properties (see Figs. 1C and 1D).

The total data containing different atomic configurations is divided into three sets including: training, validation, and test data sets. The difference between the VCNN-predicted results and true labels (i.e., formation energy or vacancy migration energy) calculated with atomistic simulations using an empirical interatomic potential constitutes the cost function. The cost is minimized during the learning process on the training data set by updating the weights and biases in the CONV and FC layers using backpropagation. Adam optimization was used to minimize the cost[6]. The CNN architecture design includes several hyperparameters, such as the numbers of CONV and FC layers in the architecture. Table S1 reports the hyperparameters were parametrically studied in our work. The optimum ML architecture and hyperparameters were chosen based on the best performance in the validation data set by cross-validation and then this architecture was used to predict the errors and properties of the test data set.

To analyze the performance of our framework, we conducted two case studies on the Ni-Al material system, which is widely used in high temperature applications such as jet engine turbine blades[7]. $Ni_{1-x}Al_x$ with $x < 30\%$ has chemically disordered fcc and ordered $L1_2$ phases. Choice of this model system enabled us to use a reliable atomistic potential to produce the true labels for our training and validation data sets[8]. For the first case study, the periodic simulation box consisted of 4×4×4 fcc unit-cells, with 256 atoms in total. Each atom can be Al or Ni, subject to a maximum of 30% Al in total. Therefore, there exist $2^{77} \sim 10^{23}$ brute force configurations in this space. We randomly generated 200,000 configurations with different Al %, as shown in Fig. 2A.

We encoded an 8×8×8 tensor into the CNN architecture. The training, validation, and test data constitute 60%, 20%, and 20% of the total data. The true labels for the training and validation data were obtained using minimization (see Methods). The hyperparameters of the architecture which were optimized by cross-validation are reported in Table S1. The predicted formation energies with respect to calculated values are presented in the inset of Fig. 2B, showing an excellent match. We also analyzed the prediction mean absolute error (MAE) for different training data percentages. As shown in Fig. 2B, the normalized prediction error of the test data decreases to 1.0% (the error divided by the range of the change of the variable) as 5% of the total data is used as a training data. The normalized prediction error for crystal graph CNNs with respect to density functional theory (DFT) calculations for 20% of the data was reported to be 1.3%.[9] These two errors are in the same order of magnitude. Moreover, the *training* root mean square error (RMSE) of cluster expansion for a binary alloy Al-Li was reported as 5 meV/atom[10]. Our RMSE for the *test* data is about half the reported MAE in Fig. 2B (0.9 meV/atom for 5% of training data). These comparisons illustrate how VCNN can accurately and efficiently predict a material property for different compositions in varied phases.

In Fig. 2C we present the relationship between the prediction MAE of VCNN and the depth of the architecture. Two architectures were analyzed. In the first one, the input tensor is connected directly to one, two, three, or four FC layers; we refer to this as the VNN. In the second architecture, the input tensor initially passes a CONV layer, and then passes one to four FC layers. We find that the presence of an initial CONV layer for the shallowest architecture (1 FC layer of 1 neuron) decreases the prediction MAE from ~11 meV/atom to 2.7 meV/atom. This is because, similar to image analysis, the CONV layers, through kernel filters, contribute to detect rotational and translational symmetries and to extract critical features from the input tensor. Moreover, the CONV layers avoid the overfitting limitation of FC layers. Due to the combined feature extraction and inherent nonlinear optimization capability of the VCNN, its accuracy surpasses the VNN.

Including two FC layers in the architecture decreases the prediction error significantly compared to one layer. However, further FC layers do not contribute notably to lower prediction error. The optimum architecture depth should correlate directly with the degree of complexity due to the number of species (here, alloying elements) in the model and the strong correlations (complexation/speciation) that may form out of them. As shown in Fig. S2, the computational cost does not follow a specific trend by adding FC layers. Thus, the optimum architecture is chosen as that with maximum prediction accuracy in the present study.

Next, we performed semi-grand Monte Carlo simulations[11] with this VCNN on-lattice potential. As shown in Fig. 2D, we predicted a single axis of the Ni-Al phase diagram at 600 K. Al remains a random solid solution in the Ni fcc crystal with increasing concentration up to 15%. Further increase of the relative chemical potential of Ni and Al shows an abrupt jump in Al%, indicating formation of the L1$_2$ phase at 25% of Al. Therefore, the VCNN predicts the phase diagram with the same accuracy as the original interatomic potential[8].

In the second example, we studied the energy landscape of vacancy diffusion in the Ni$_{1-x}$Al$_x$ model system, at $x$ = 15 at%, corresponding to the maximum Al% in the fcc phase. Substitutional diffusion is mediated by exchanging sites of Ni and Al atoms with a vacancy. The local arrangement of these atoms around a vacancy influences the associated energies for diffusion. The representative atomistic model (Fig. 3A) consists of 2×2×2 unit cells of the fcc crystal, and the boundary conditions are periodic in all three dimensions. A single vacancy was placed in the center site. The fcc crystal has 12 and 6 atoms on the first and second nearest neighbor shells of the vacancy, respectively (see Fig. S3A). The energies associated with the site exchange are mainly influenced by the first nearest neighbor atoms to the vacancy. For simplicity, we thus investigated only the influence of the presence of Al atoms in these sites on the target energies. The rest of the atoms are Ni. Hence, Al atoms constitute 4 out of the 12 first nearest neighbor atoms. The total number of possible configurations of these 4 atoms around the vacancy is $\binom{12}{4} = 495$. Considering 12 possible exchanging sites for each configuration yields in total 5,940 final configurations for this model.

After generating all initial and final configurations, i.e., pair configurations before and after exchanging sites with vacancies, 60%, 20%, and 20% of them were assigned as training, validation, and testing data sets, respectively. The procedural diagram is in Fig. S4. As shown in Fig. 3A, for each atomic exchange, the VCNN regressor can predict four quantities: the initial and final formation energies ($E_1$, $E_2$) and the forward and backward energy barriers ($\Delta E_1$, $\Delta E_2$). Three of these four energies are independent variables. The training and validation configurations were relaxed using minimization. Two labels, $E_1$ and $E_2$, are thus obtained by the relaxation of the configurations. In the next step, each pair of the relaxed initial and final configurations is fed into a climbing image nudged elastic band (CINEB) calculation to obtain the labels for the migration energy barriers. The details of these atomistic simulations are presented in the Methods section.

The next step is to encode the atomistic models into the VCNN architecture. Our approach is similar to the previous example, but now it is necessary to encode the additional vacancy site as an empty site in the input tensor. The total number of input variables is 27 (Fig. S3B). The tensor of the final configuration was encoded as our input data.

To proceed, we used the same VCNN architectures as in the first case study (see Table S1). We also optimized the hyperparameters for the VCNN architectures by cross-validation. The optimum architecture was the same in both case studies for all hyperparameters, except the kernel size of CONV layers (5 and 3 for case studies one and two, respectively). Therefore, the optimal VCNN architecture is not dependent on the input size data. The inset in Fig. 3B presents the predicted and calculated $\Delta E_1$, which shows an excellent match between the two values. The MAE on the test sets of $\Delta E_1$, $\Delta E_2$, and also $E_2$ with respect to the training data percentage are plotted in Figs. 3C and S5. For the most accurate VCNN architecture, the MAE of $\Delta E_1$ and $E_2$ decreases below ~20 meV and 1 meV/atom, respectively, when > 40% of the total data is used as the training data set.

The predicted $\Delta E_1$, $\Delta E_2$, as well as $E_2$, are shown in Figs. 3C, 4A, and S6A. These energies can then be fed to kinetic Monte Carlo simulations for calculating the diffusivity of each of the alloying components. The distribution of energies can provide guidelines on the components' diffusivity in the studied alloy system. For example, $\Delta E_1$ in this model system has a bi-step distribution (Fig. 3C). The step with a higher magnitude of energy barrier (> 1.1 eV) is attributed to the configurations wherein Al exchanges its site with a vacancy (see the violin plot in Fig. 3D). All low energy barriers are related to the diffusion of Ni atoms, and thus Al atoms should be less prone to diffusion if this energy alone determines diffusion. The $\Delta E_2$ has a Gaussian distribution (see Fig. S6A). The classification of this energy for Ni and Al atoms reveals that all extreme, maximum, and minimum values are related to the diffusion of Al atoms. Diffusion of Ni atoms constitute the core of the normal distribution (see Fig. S6B). Seven distinct steps can also be captured in the $E_2$ distribution (see Fig. 4A). Classifying this energy also for Ni and Al atoms shows that, on average, the configurations with mobile Al atoms achieve higher energy levels than the ones of Ni atoms (see Fig. 4B). Therefore, by analyzing the distribution of these three energies, we expect that Al atoms should have lower diffusivity in the Ni matrix.

Last, we analyzed how the atomic arrangement of Al around the vacancy influences $E_2$. To quantify the arrangement (clustering), we applied a geometrical parameter that is frequently used to estimate the average of the angle between ternary pairs of atoms in a cluster[12,13] (see Fig. 4C).

$$\Psi(i;\xi,\lambda,\zeta) = \sum_j \sum_k e^{-\frac{R_{ij}^2 + R_{ik}^2 + R_{jk}^2}{\xi^2}} (1 + \lambda \cos \theta_{ijk})^\zeta, \qquad 1$$

where $R_{ij}$ is the distance between atoms i and j, $\theta_{ijk}$ is the angle between atoms $i, j$, and $k$ (see Fig. 4C). $L, \mu, \xi, \zeta$, and $\lambda$ are constants (see Table S2). This parametrization ensures that values of $\Psi$ differentiate distinct configurations of Al atoms, while capturing rotationally and translationally symmetric configurations as equivalent[13]. For instance, when four Al solutes are present in the first nearest neighbors of a vacancy, owing to translational and rotational symmetries, there are only 14 values of $\Psi$ in unrelaxed initial configurations. The value of $\Psi$ increases as the Al atoms become nearer to one another and develop SRO, as shown in the Fig. 4D insets. Fig. S7 shows a violin plot of $E_2$ classified based on these values ($\Psi_1$-$\Psi_{14}$). The average $E_2$ (black circle points) increases by ~0.03 eV/atom as $\Psi$ increases. Further analysis of the bimodal distribution of the violin plot for all $\Psi$ reveals that all cases with high $E_2$ are related to the Al atoms exchanging sites, and all low energies are related to the Ni atoms. We then replotted the distribution by separating $E_2$ for Al and Ni, in Figs. 4D and 4E, respectively. $E_2$ increases by 0.015 eV for configurations with mobile Al atoms (see Fig. 4D). Figure 4E shows that $E_2$ for configurations with mobile Ni rises ~0.03 eV/atom as $\Psi$ increases, which indicates SRO of Al atoms. Therefore, we predict that a random distribution of Al atoms in the first nearest neighbor of a vacancy should enhance the diffusivity of Ni atoms in Ni-Al. More interestingly, five distinct steps are seen in the distribution of $E_2$ for configurations with mobile Ni atoms (see Fig. 4B), and these five steps are also present when the data is classified based on $\Psi$ (see Fig. 4E). These analyses show that to increase the diffusivity of Ni in Ni-Al, one needs to decrease the segregation of Al atoms. This decreased segregation reduces $E_2$ and thus increases the probability of the Ni diffusion. The benefits of extracting atomic-scale SRO thus include defining the correlation energies and revealing how the atomic-scale local arrangement can strongly influence global energies of the material.

The on-lattice VCNN has several advantageous compared to off-lattice potentials such as NN[12] and crystal graph CNN[9]. For the case of NN potentials, initially symmetry-invariant transformation features need to be specified manually. The size of the vector also increases significantly by increasing the number of alloying elements similar to the clustering method. In case of crystal graph CNNs, the links between the interacting atoms must be defined a priori, which is especially numerically expensive for close packed crystals which are practically relevant to many alloys. Therefore, VCNN presents a relatively low computational cost compared to NN and CNN approaches.

Going forward, VCNN potentials can be applied to study complex-concentrated alloys such as high/medium entropy alloys and superalloys. These alloys can exhibit an exceptional combination of superior damage tolerance and strength at extreme temperatures[7,14]. In addition to the immense compositional breadth, their alloying elements follow a particular ordering in the first few neighbor atomic shells on the scale of 1-3 nm and form SROing[15–17]. Therefore, due to compositional complexity, neither single-phase stabilization through tailoring of elemental diffusivity nor crystallographic defect interaction (e.g., dislocations), which lead to their unique deformation mechanisms[17,18] have been systematically investigated and confirmed. Atomistic simulations provide a reliable platform for systematic study of these phenomena; however, production of a reliable and efficient potential was the bottleneck for simulations, which now can be relieved by VCNN potentials.

In conclusion, we presented the VCNN approach as a new on-lattice potential, as an alternative to the established cluster expansion method. The VCNN encodes the positions of the atoms directly as input, and thus it does not require manual preprocessing of the input data. Via two case studies on the Ni-Al model system, we demonstrated the generality of our framework for the prediction of different material properties for crystals with various phases, compositions, and defect locations. Analysis of the relationships between the configurational energies and atomic positions enables understanding of how atomic ordering influences global material properties such as energies associated to diffusivity. In future work, other atomistic properties such as electronegativity, magnetic moment, and atomic size (Fig. 1B) can be also encoded into the ML architecture as additional 'color' channels. The VCNN framework can therefore be readily applied to study the dependence of various properties on atomic arrangement in diverse multi-component alloys, without additional preprocessing cost.

**Acknowledgments**

S.M.T.M., R.W.P., and A.J.H. acknowledge financial support from ArcelorMittal S.A., the MIT-Portugal Program (MPP2030), and a gift from Robert Bosch, LLC. J.L. and S.S.M.A. acknowledges support by NSF CMMI-1922206 and Timken. The authors acknowledge fruitful discussions with Shaolou Wei, Qing-Jie Li, and Nigamaa Nayakanti. The authors also acknowledge the MIT Supercloud and Lincoln Laboratory Supercomputing Center for providing (HPC and consultation) resources that have contributed to the numerical results reported within this paper.

**Author contributions**

S.M.T.M. conceived and designed the voxelated convolutional neural network, wrote the codes and conducted atomistic simulations, machine learning optimizations, and postprocessing. S.M.T.M., S.S.M.A., and J.L. designed the atomistic models, and S.M.T.M. and J.L. analyzed the results. S.M.T.M., A.J.H., and J.L. designed the methodology and wrote the paper. All authors participated in discussions and commented on the manuscript.

**Competing interests**

Authors declare no Competing Financial and Non-Financial Interests.

**Methods**

**Minimization**

The simulation cell in the first case study consists of 64 fcc unit cells. The model has 256 atoms, in which Al constitutes 30% of atoms at maximum. For the second case study, eight fcc unit cells with a vacancy in the center site. The model thus contains 31 atoms. Al atoms are present in the first nearest neighbor surrounding the vacancy and constitute 4 out of these 12 atoms. The rest of the atoms are Ni. While the presence of Al atoms in the second nearest neighbor may also influence the energies associated with the exchanging sites with vacancies, for simplicity and to show the proof-of-concept of the model these interactions were not considered in this study. Periodic boundary conditions were imposed at all boundaries of the sample. The embedded atom method potential was adapted for Ni-Ni, Ni-Al, and Al-Al interatomic interactions[8]. The atomic configurations before and after exchanging sites with the vacancy was minimized using conjugate gradient technique in lammps[19], with the tolerance energy and force errors of $10^{-18}$.

**CINEB calculation**

The energy barrier for exchanging sites with the vacancy was quantified using reaction path calculations. We used the climbing image nudged elastic band (CINEB) method[20], implemented in lammps[19] to determine the minimum energy path (MEP) of a reaction (exchanging of a site of atoms with a vacancy). The activation energy was given by the maximum on the MEP, a saddle point on the potential energy surface of the system. The relaxed configurations before and after exchanging sites with the vacancy were used as initial and final states in the CINEB calculation. The error tolerances for force and energy were chosen to be $10^{-5}$.

**ML approach**

The ML calculations were performed using Tensorflow[21] running on 16 Nvidia Volta V100 GPU cores (first case study) and a single-core CPU (second case study). The reported calculation time (see Fig. S2) was reported from the analysis on 16 GPU cores. Deep learning methods with various architectures were assessed (see Fig. 1, and Table S1). The weights were initialized randomly and then added with biases and the rectified linear unit activation function was applied on them. The hyperparameters were chosen as $10^{-3}$ for learning rate, 150,000 for the number of training loops, 128 for the batch size, $L_1 = 0$, $L_2 = 0$, and 5×5×5 and 3×3×3 as convolutional kernel size for case study one and two, respectively, if they were not changed for the parametric studies. The MAE of prediction was measured for the test data for each training percentage and each simulation was repeated five times for five different random initial variables.

**Data availability**

All data required to reproduce the findings during this study are included in this Manuscript and Supplementary Information.

**Main figures:**

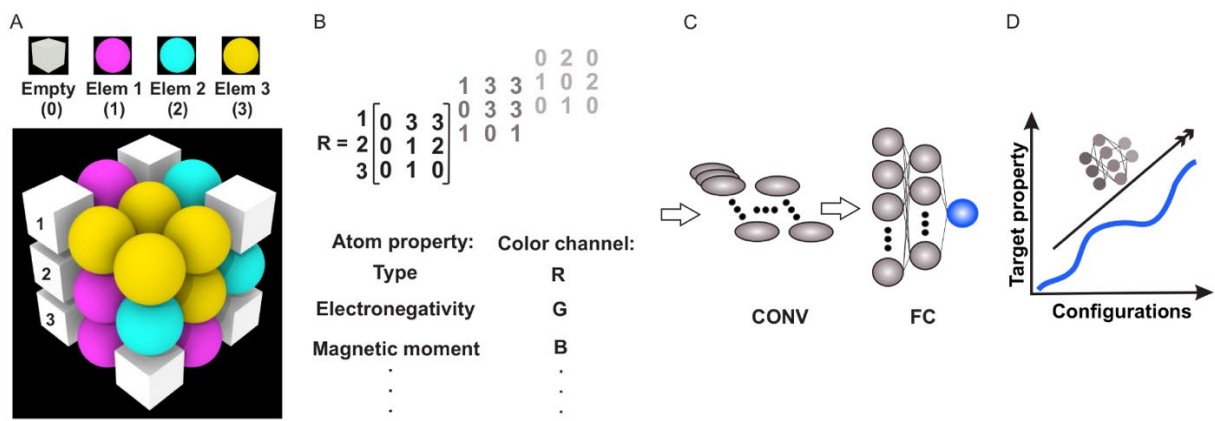

Fig. 1. The voxelated convolutional neural network (VCNN) framework. A) Indexing the occupied and empty voxels in the simulation box. B) Encoding the indexed voxels into a tensor to be fed to the ML architecture. Different color channels such as atom type, electronegativity, and magnetic moment can be considered in the input data. C) The ML architecture, which is a combination of convolutional (CONV) and fully connected (FC) layers contains several CONV features and neurons, respectively. The target properties to be predicted by the trained ML algorithm are stored in the neuron of the last FC layer. D) The predicted properties can be optimized by exploring the design space by the ML algorithm.

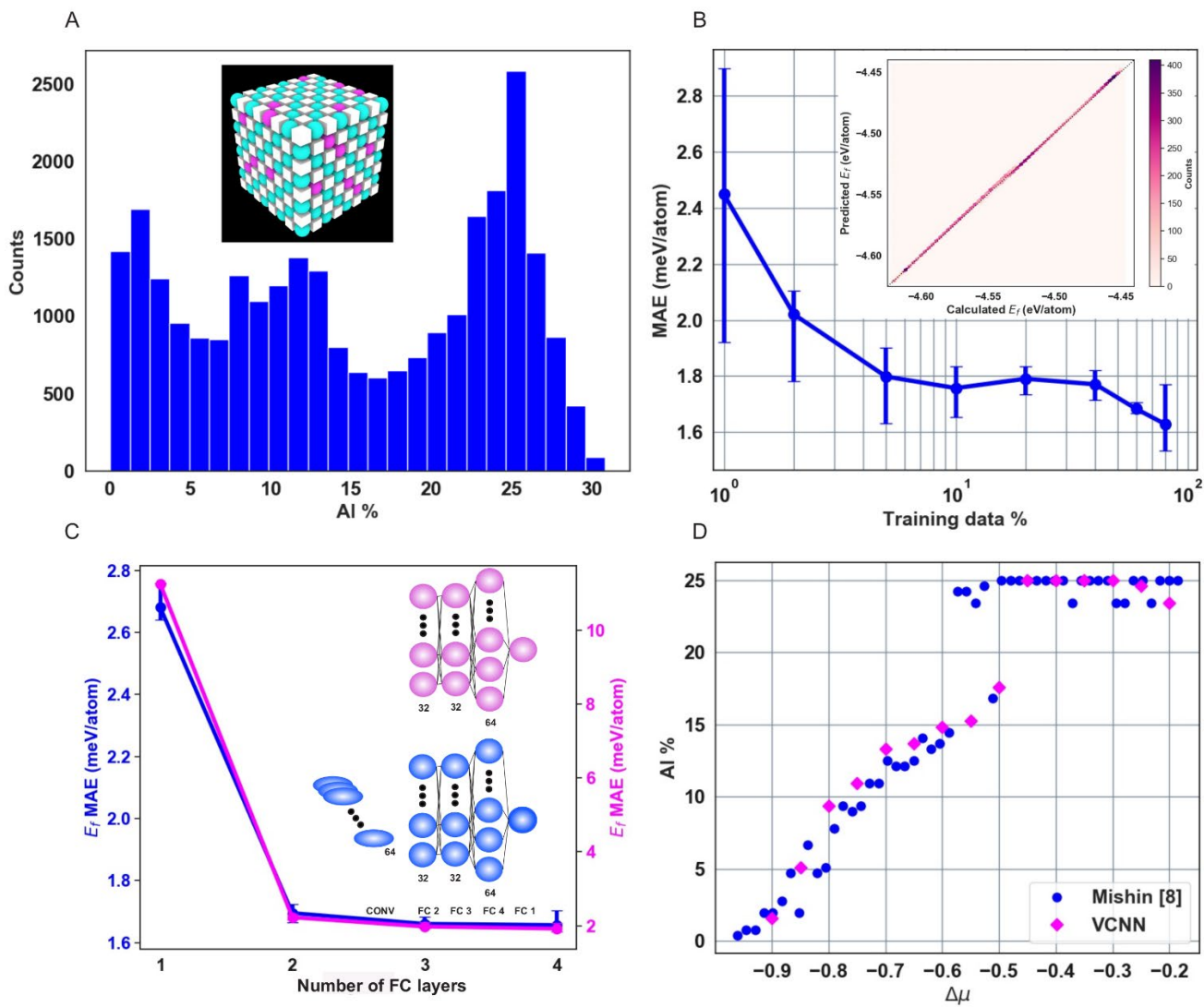

Fig. 2. A) Distribution of Al % in the input data set. The inset shows the input model of Ni-Al. The sites in the cubic crystal are colored differently (Al: magenta, Ni: cyan, empty sites: white). B) The mean-absolute error (MAE) of the test data with respect to the training data percentage for the formation energy. The inset shows the excellent match between the predicted and calculated formation energy $E_f$ of the test data set. C) The prediction error with respect to the number of FC layers in two studied architectures. D) 600 K single axis in Ni-Al phase diagram. Each point represents the Al% of a fully converged semi-grand Monte Carlo simulation for a given relative chemical potential of Ni and Al ($\Delta\mu$).

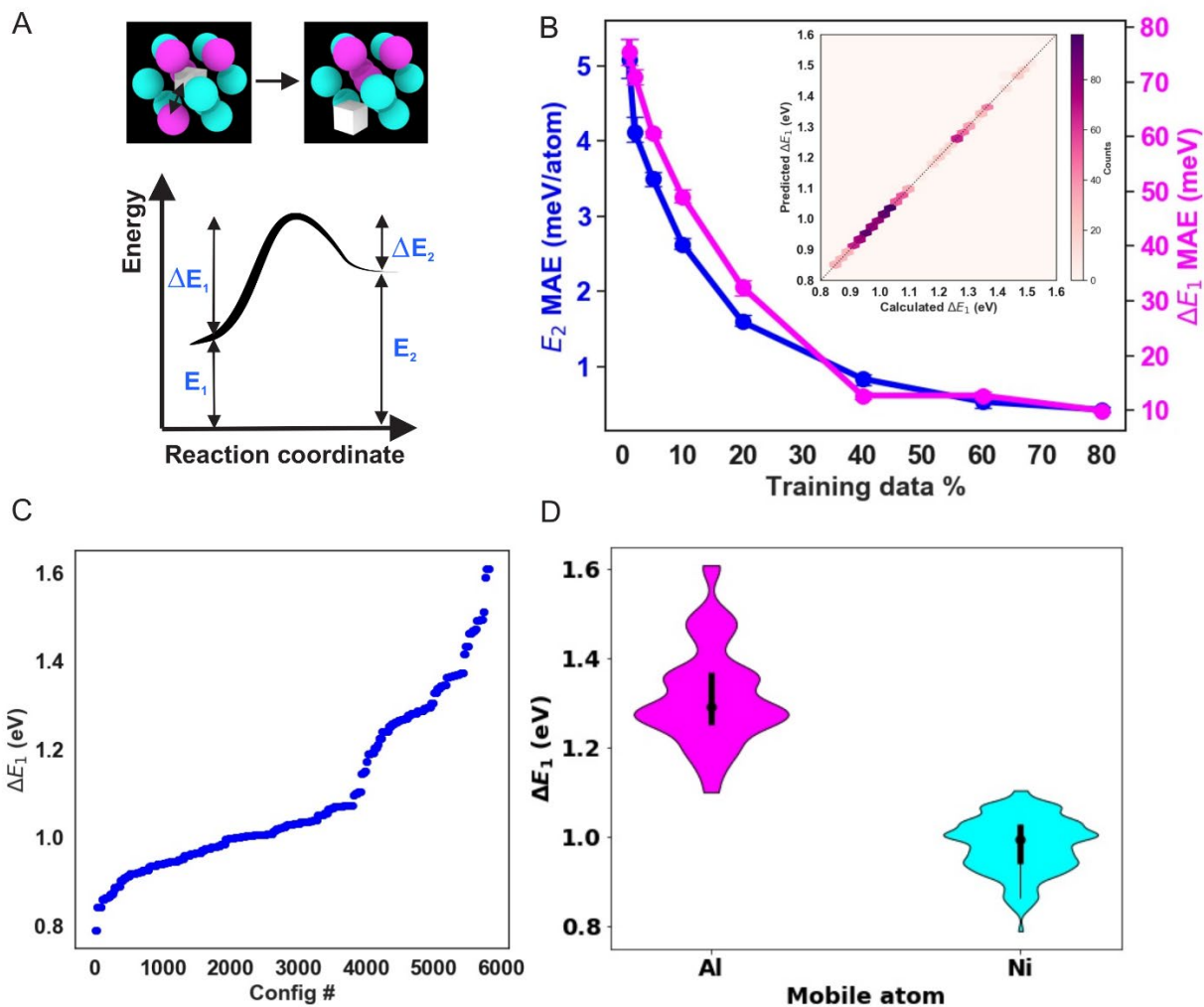

Fig. 3. A) 12 first nearest neighbor of the vacancy in an fcc crystal of Ni (cyan color) with 4 Al atoms (magenta color). The target energies to be predicted by trained ML algorithm include: forward and backward energy barriers ($\Delta E_1$, $\Delta E_2$), and initial and final formation energies ($E_1$, $E_2$). B) The mean-absolute error (MAE) of the test data with respect to the training data percentage for $\Delta E_1$ and $E_2$. The inset shows the excellent match between the predicted and calculated forward energy barrier $\Delta E_1$ for the test data set. C) Predicted $\Delta E_1$ by ML algorithm for all configurations. D) The violin plot presents the classification of $\Delta E_1$ based on the type of atom (Ni/Al) which is exchanging sites with a vacancy.

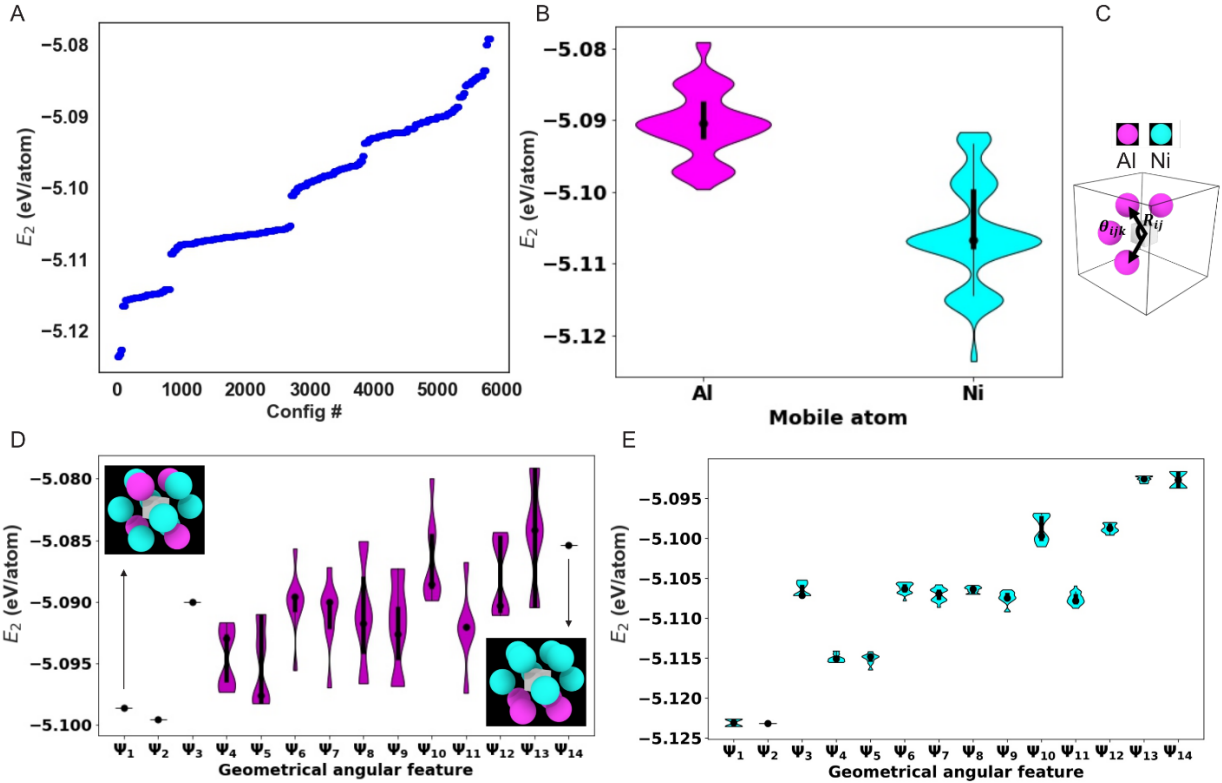

Fig. 4. A) Predicted final formation energy $E_2$ by VCNN algorithm for all configurations. B) The violin plot presents the classification of $E_2$ based on the type of atom (Ni/Al) which is exchanging sites with a vacancy. C) Schematic showing the distance and angular parameters used in Eq. 1. The data is further classified based on the type of atom which is exchanging sites with a vacancy: D) Ni, E) Al.

# SI figures & tables:

| Hyperparameter | Range |
| --- | --- |
| CONV feature # | 64, 64, 128, 128, 64, 64 |
| FC neuron # | 32, 32, 64, 64 |
| CONV kernel size | 2, 3, 4, 5, 6, 7, 8 |
| # of FC layers | 2, 3, 4 |
| # of CONV layers | 1, 2, 4, 6 |
| Batch size | 32, 64, 128, 256, 512, 1024, 2048 |
| Training loop number | 100, 1000, 10,000, 150,000, 200,000 |
| Learning rate | $10^{-4}$, $10^{-3}$, and $10^{-2}$ |
| Regularization factor $L_1$ | $10^{-6}$, $10^{-5}$ |
| Regularization factor $L_2$ | $10^{-2}$, $10^{-3}$ |

Table S1. Hyperparameters which were parametric studied in both case studies presented in the manuscript. The optimum architecture contains one CONV layer with 64 features, two FC layers, each with 32 neurons, batch size of 128, 150,000 training loops, $10^{-3}$ learning rate, $L_1 = 0$, $L_2 = 0$, and CONV kernel size of 5 and 3 for case studies one and two, respectively.

| *L* | *μ* | *ξ* | *ζ* | *λ* |
| --- | --- | --- | --- | --- |
| 1 | 0 | 7.6 | 1 | 1 |

Table S2. Constant variables in Eq. 1.

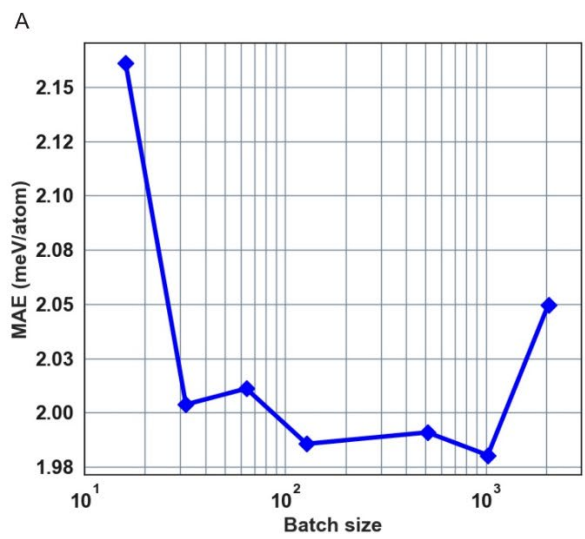 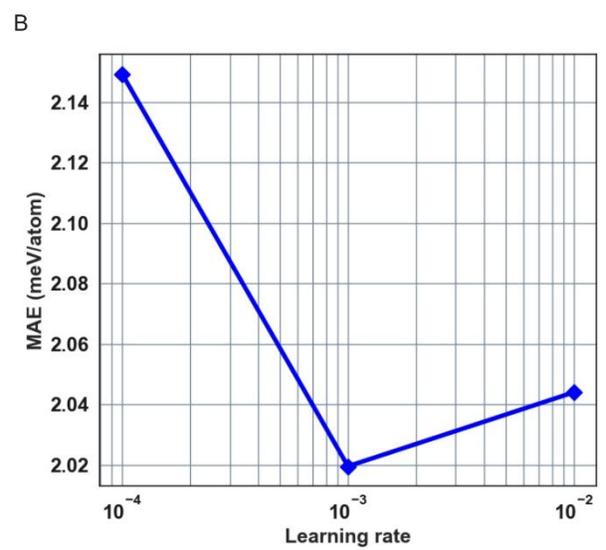

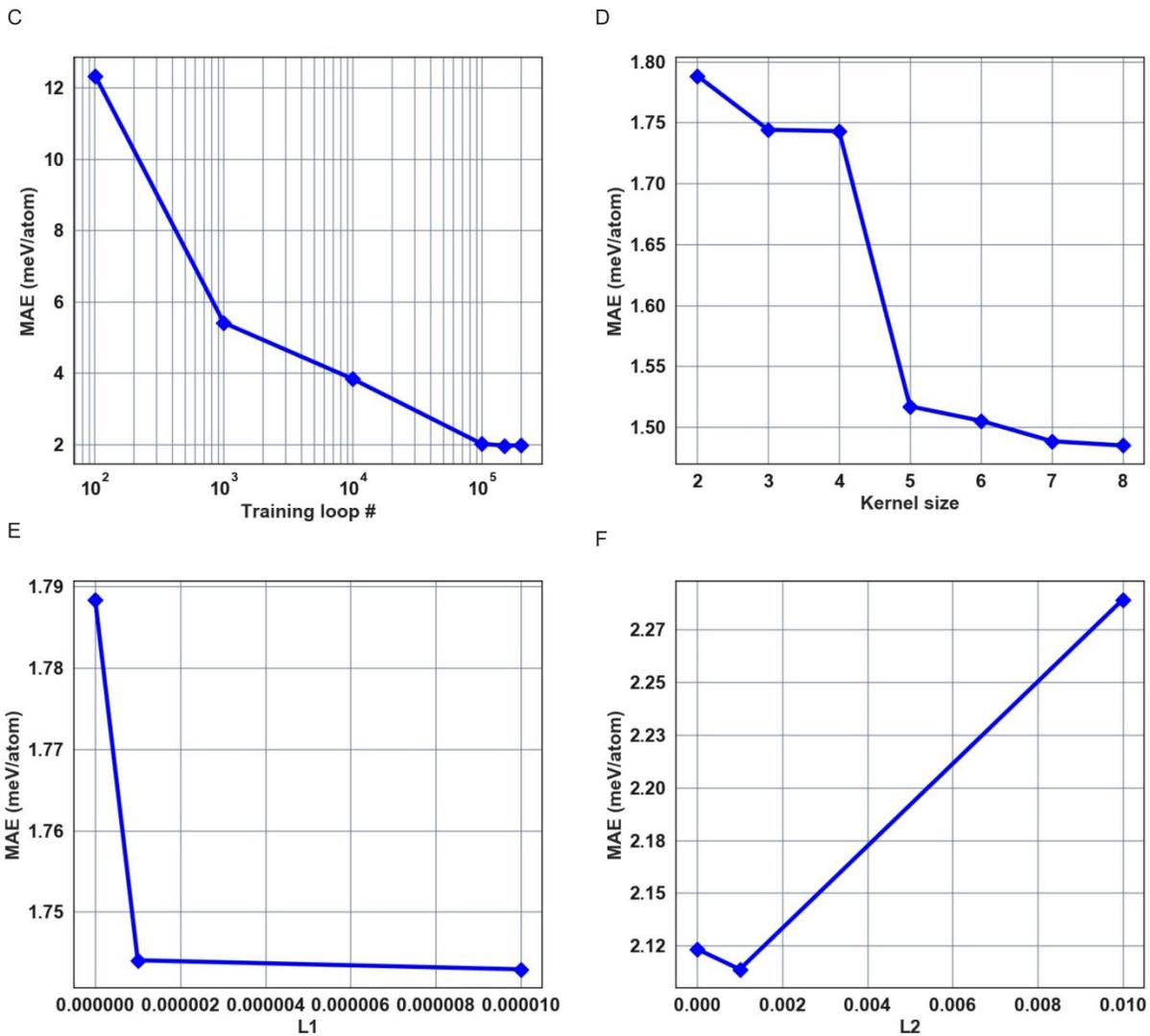

Fig. S1. Parametric studies on ML hyper parameters. MAE of prediction for different A) batch sizes, B) learning rates, C) number of training loops, D) kernel sizes, and E,F) $L_1$ and $L_2$ regularization parameters.

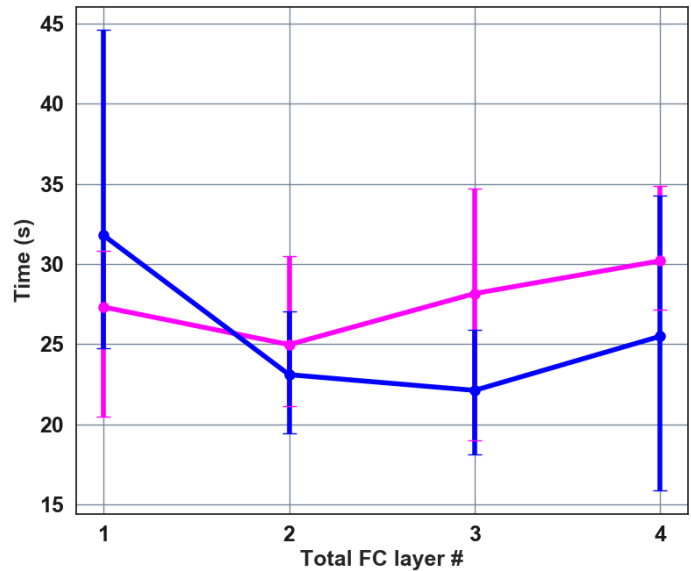

Fig. S2. Training time for different ML architectures presented in Fig. 2C, for 1000 number of training loops. The simulations were performed on 16 GPU nodes.

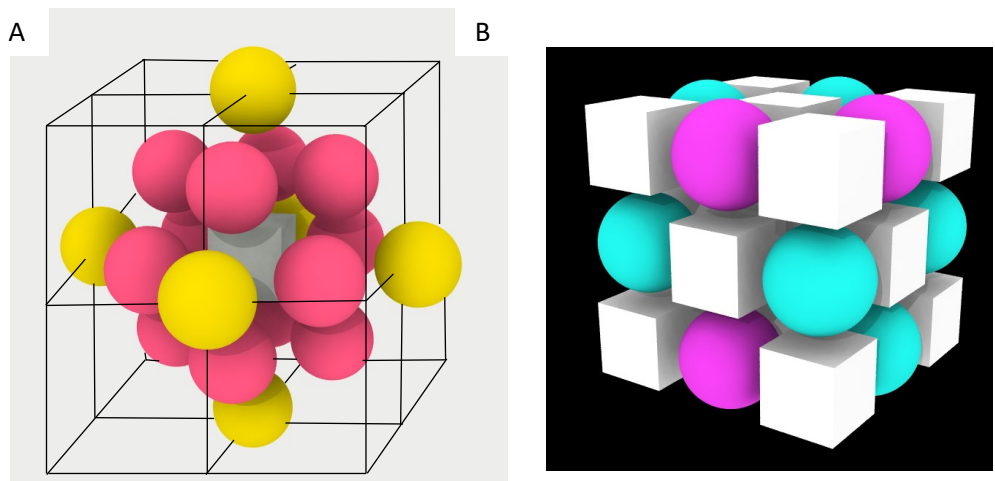

Fig. S3. A) The atomistic model containing eight unit cells of fcc crystal, and a vacancy located in the center site. Only first (pink) and second (yellow) nearest neighbor atoms have been shown in this model. B) The input model of Ni-Al for case study two. The sites in the cubic crystal are colored differently (Al: magenta, Ni: cyan, empty sites: white).

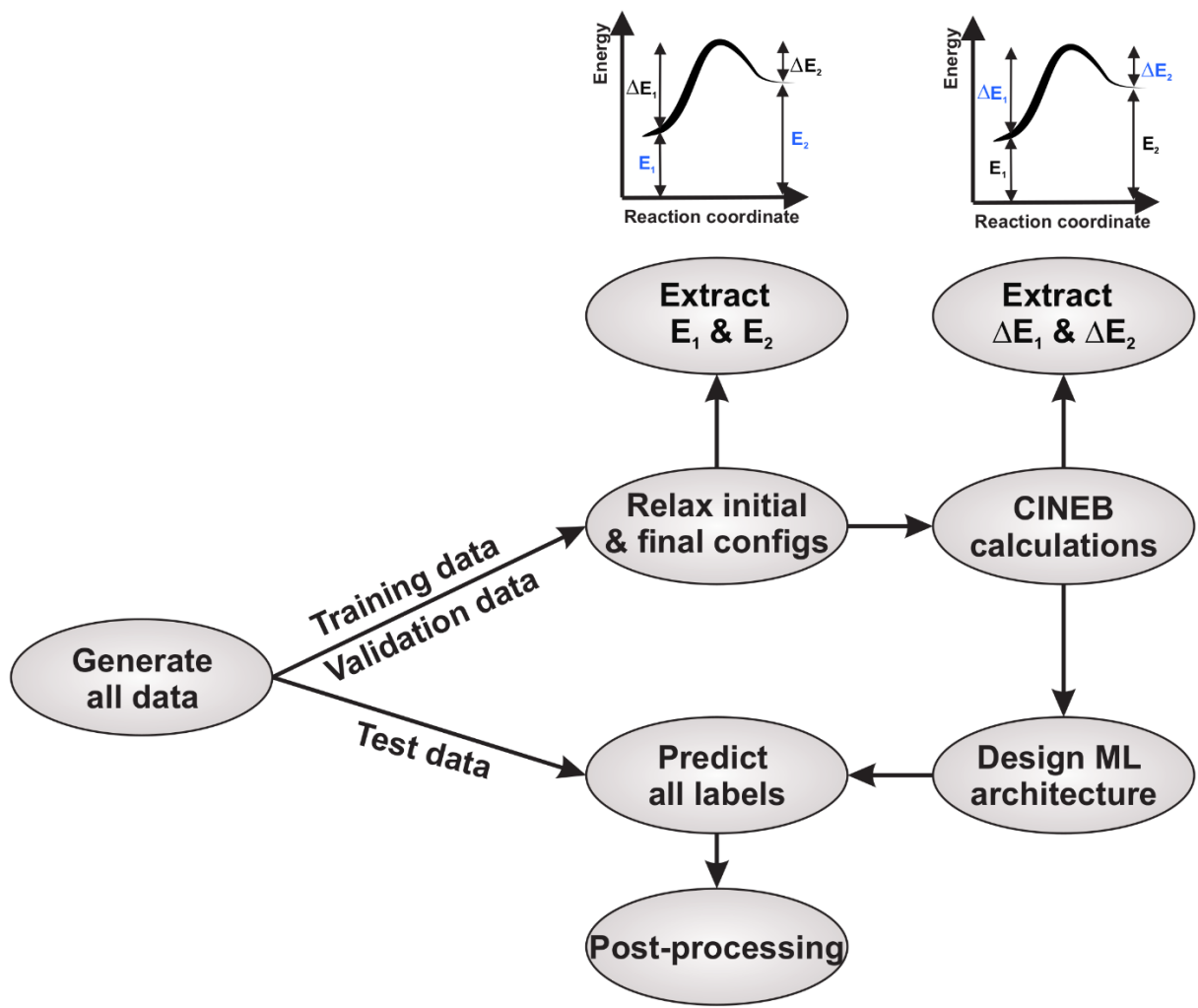

Fig. S4. Overall flow diagram of our framework for the second case study.

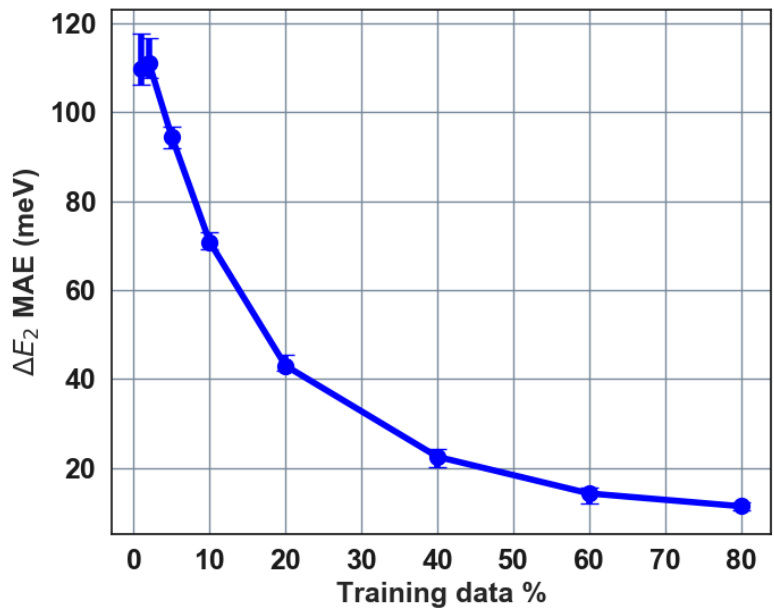

Fig. S5. The mean-absolute error (MAE) of the test data with respect to the training data percentage for the backward ($\Delta E_2$) energy barriers in the second case study.

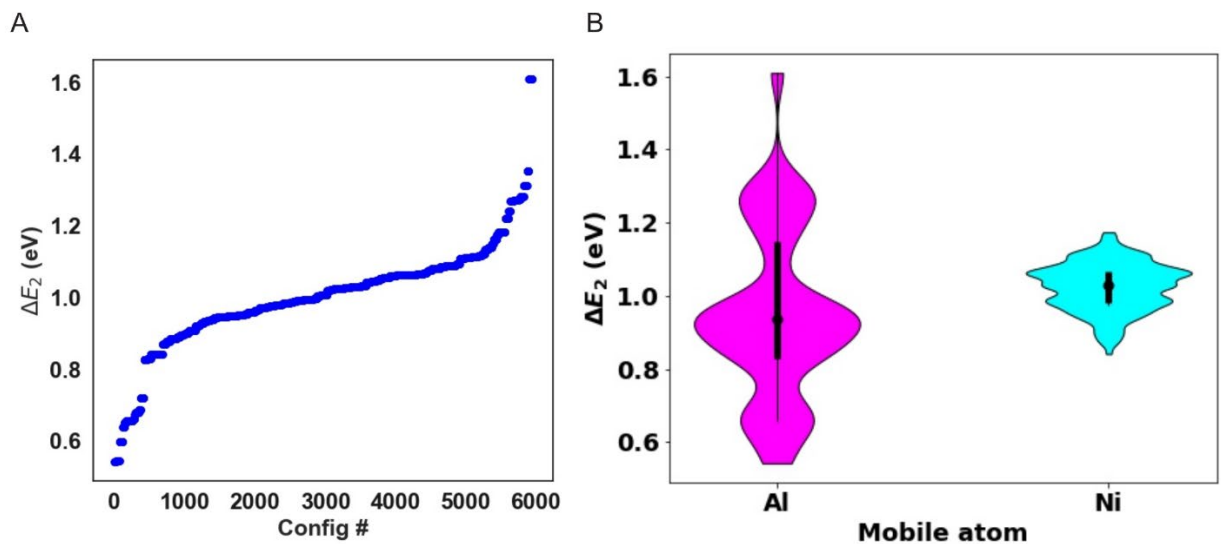

Fig. S6. Predicted backward energy barrier by the ML algorithm for all configurations. B) The violin plot presents the classification of this energy based on the type of atom (Ni/Al) which is exchanging sites with a vacancy.

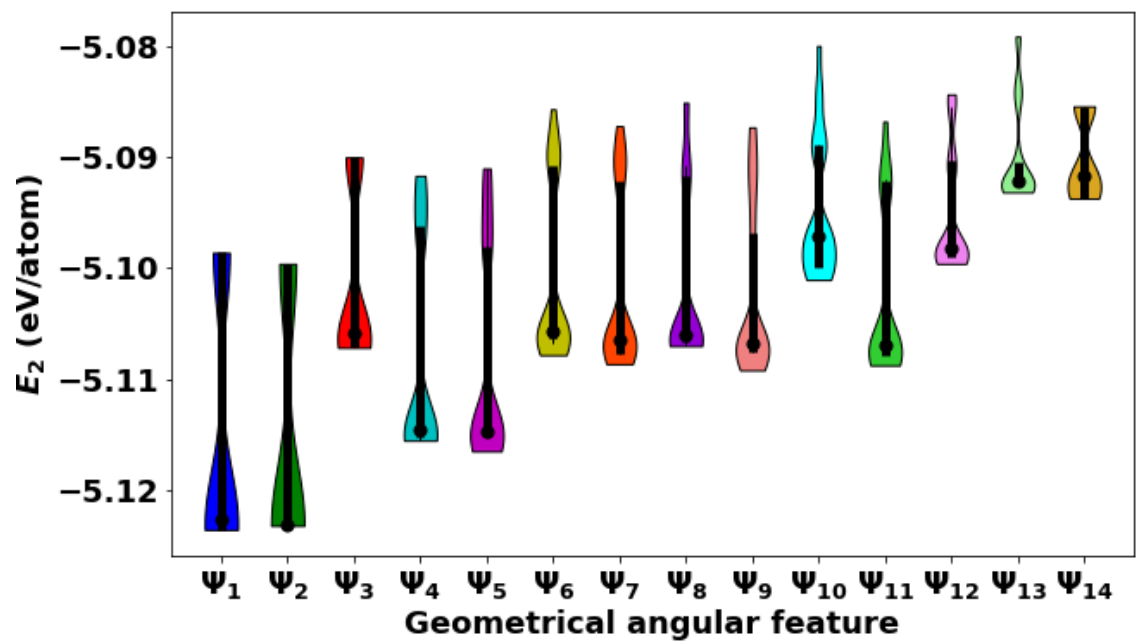

Fig. S7. Classifying the final formation energy based on the geometrical angular feature. The violin plot indicates bimodal distribution for each Ψ.